\documentclass[12pt]{article}
\usepackage{amsfonts,amsbsy}
\newcommand{\be}{\begin{equation}}
\newcommand{\ee}{\end{equation}}
\newcommand{\ba}{\begin{array}}
\newcommand{\ea}{\end{array}}
\newcommand{\p}{\partial}

\DeclareMathAlphabet{\bi}{OML}{cmm}{b}{it}

\def\id{\mathop{\rm id}\nolimits}

\newcommand{\fl}{\hspace*{-5mm}}

\date{}

\author{M. Marvan and A. Sergyeyev\\
Mathematical Institute,
Silesian University in Opava,\\ Bezru\v{c}ovo n\'am. 13, 746\,01 Opava,\\
Czech Republic\\ 
E-mail: {\tt Michal.Marvan@math.slu.cz},
\tt{Artur.Sergyeyev@math.slu.cz}}
\begin{document}
\title{\bf Recursion operator for stationary Nizhnik--Veselov--Novikov 
equation}
\maketitle

\begin{abstract}
We present  a new general construction of recursion operator from 
zero curvature representation. Using it, we find
a recursion operator for the stationary Nizh\-nik--Ve\-se\-lov--Novikov
equation and present a few low order symmetries generated with the help
of this  operator.
\looseness=-1
\end{abstract}



In the present letter we suggest a new method of construction of 
recursion operator using the zero curvature representation.
Unlike the majority of the hitherto known methods, 
see, e.g., 
\cite{akns}--\cite{skg}
and references therein, 
ours is immediately applicabile 
to both evolutionary and non-evolutionary systems,
and gives not only the recursion operator, but also its inverse, 
leading to the `negative' part of the hierarchy of the system in
question. We apply the method to the stationary 
Nizhnik--Veselov--Novikov
(NVN) equation for which no recursion operator 
has been found so far.
\looseness=-1

Let $F=0$ be a system of PDEs in two independent variables $x,y$ for 
the un\-known $n$-com\-ponent vector function
$\bi{u}=(u^1,\dots,u^n)^{T}$,  where the superscript `$T$' 
denotes matrix transposition.  Let this system have 
a zero curvature representation
$D_y (A)-D_x(B)+[A,B]=0$, where 
$A$ and $B$ take values in a (matrix)
Lie algebra $\mathfrak{g}$ and depend on $\lambda, x,y$ and 
$\bi{u}$ and its
derivatives. 
Here $D_x,D_y$ are the operators of total $x$- and $y$-derivatives, 
see, e.g., Ch~2 of \cite{olv_eng2}, and \cite{vinbook}.
Note that $A$ and $B$ may involve an essential (spectral) 
parameter $\lambda$.
\looseness=-2

Consider a (possibly vector or matrix) function
$P$ of $x,y$, $\bi{u}$ and its derivatives. Then 
the directional derivative of $P$ along an $n$-component
vector 
$\bi{U}=(U^1,\dots,U^n)^{T}$
is given by $P'[\bi{U}]=\sum_{\alpha=1}^{n}\sum_{i,j=0}^{\infty}(\p P/\p
u^\alpha_{ij})
D_x^{i}D_{y}^{j}(U^\alpha)$, where $u^\alpha_{00}\equiv u^\alpha$,
$u^\alpha_{ij}=\p^{i+j}u^\alpha/\p x^i\p y^j$, cf, e.g.,
\cite{bl}. 
In \cite{vinbook} 
$P'[\bi{U}]$ is called a linearization and denoted $\ell_P \bi{U}$. 
\looseness=-1

Let $\bi{U}$ be a symmetry of the system $F=0$, that is,
let $\bi{U}$ satisfy $F'[\bi{U}]=0$ on the solution manifold of $F=0$
\cite{olv_eng2, vinbook}. 
Consider a $\mathfrak{g}$-valued solution 
$S$ of the system 
\be\label{ro0}
D_x(S)-[A,S]=\tilde A\equiv A'[\bi{U}], \quad
D_y(S)-[B,S]=\tilde B\equiv B'[\bi{U}]. 
\ee
Assume that we have found $n$ linear combinations
$\tilde U^\alpha=\sum_{i,j} a^{\alpha,ij}S_{ij}$ of entries $S_{ij}$ of
$S$,
$\alpha=1,\dots,n$,
with the property that
$\tilde\bi{U}=(\tilde{U}^1,\dots,\tilde{U}^n)^{T}$ is another symmetry of
$F=0$. Then the linear operator $\mathfrak{R}_0$
defined by $\tilde\bi{U}=\mathfrak{R}_0(\bi{U})$ is a recursion 
operator, in Guthrie's \cite{gut1} sense, for the equation $F=0$.
The coefficients $a^{\alpha,ij}$ may depend on $\lambda, x,y$ and
$\bi{u}$ and its derivatives. 
\looseness=-1 

However, testing the above scheme on a number of known examples like KdV
or Harry Dym  equation shows that $\mathfrak{R}_0$ 
generates the nonlocal
symmetries that belong to 
the `negative' part of the hierarchy of $F=0$. 
Then we should, if possible, invert $\mathfrak{R}_0$ 
in order to obtain a `conventional' recursion operator $\mathfrak{R}$, 
which will generate the `positive', local part of the hierarchy in
question. 
The inversion is an algorithmic process described in~\cite{gut1}. 
Note \cite{mar} that if the coefficients of the recursion operator are
local, then so are the coefficients of its inverse.
\looseness=-1

Let us now apply this procedure to the stationary NVN equation
\begin{equation}\label{snv}
u_{yyy}=u_{xxx}-3 (vu)_x+3 (wu)_y,\quad 
w_x=u_y,\quad 
v_y=u_x.
\end{equation}
recently studied by Ferapontov~\cite{fer}, see also Rogers and Schief
\cite{rs}, in connection with isothermally asymptotic surfaces in
projective differential geometry.

The stationary NVN equation is a reduction of the NVN
equation \cite{nzh,nv}
\begin{equation}\label{nvn}
u_t= u_{xxx}-u_{yyy}-3 (vu)_x+3 (wu)_y,\quad
w_x=u_y,\quad
v_y=u_x,
\end{equation}
obtained upon assuming that $u,v,w$ are independent of $t$.
The latter 
is well known to be integrable via the inverse
scattering  transform, as it has the Lax pair 
\begin{equation}\label{lax}
\psi_{xy}=u\psi,\quad 
\psi_t=\psi_{xxx}- \psi_{yyy}-3 v\psi_x+3 w\psi_y.
\end{equation}
The NVN equation (\ref{nvn}) is the first member of the hierarchy 
describing the deformations preserving the zero energy
level of two-dimensional Schr\"odinger operator \cite{nv}.
It also naturally arises in the theory of surfaces, see \cite{rs}
and references therein, and its modified version appears in the string
theory
\cite{kon2, yamag}.

Upon 
setting \cite{rs}
$\psi=\tilde\psi\exp(\lambda t)$, where $\lambda$ is a constant,
the Lax pair (\ref{lax})
can be transformed into a zero-curvature representation for
(\ref{snv}) of the form $D_y (A)-D_x(B)+[A, B]=0$.
This representation involves an essential parameter $\lambda$, and
the matrices $A$ and $B$ belong to the semisimple Lie algebra $sl_6$ 
of traceless $6\times 6$ matrices. 
They read 
\be\label{zcr}
\fl 
A=\left(\begin{array}{cccccc}
0 & 1 & 0 & 0 & 0 & 0\\ 
0 & 0 & 0 & 0 & 1 & 0\\ 
u & 0 & 0 & 0 & 0 & 0\\ 
A_{41} & \lambda & u_y & 0 & 0 & -u\\ 
0 & 3v & 0 & -1 & 0 & 0\\
u_y & 0 & u & 0 & 0 & 0
\end{array}\right), \quad
B=\left(\begin{array}{cccccc}
0 & 0 & 1 & 0 & 0 & 0\\ 
u & 0 & 0 & 0 & 0 & 0\\ 
0 & 0 & 0 & 0 & 0 & 1\\ 
B_{41} & u_x & 0 & 0 & -u & 0\\ 
u_x & u & 0 & 0 & 0 & 0\\
\lambda & 0 & 3w & -1 & 0 & 0
\end{array}\right),
\ee 
matrices with zero trace, and
where $A_{41}=-u_{yy}+3wu$, $B_{41}=-u_{xx}+3vu$.
\looseness=-1

Let
$\bi{U}=(U,V,W)^{T}$ be a symmetry of (\ref{snv}),
i.e.\ let $U,V,W$ satisfy 
\looseness=-1
\begin{equation}\label{lin}
\begin{array}{l}
\fl 
D_{y}^3 U =D_{x}^3 U +3\left[ wD_y U+ u (D_y W- D_x V)- u_x V+ u_y W 
+(w_y- v_x)U - v D_x U\right],\\  
D_x W=D_y U, \ \ 
D_y V=D_x U.
\end{array}
\end{equation} 
  
Consider a traceless $6\times 6$ matrix $S$ 
that solves (\ref{ro0}), where $A,
B$  are given by (\ref{zcr}) and 
\[
\fl \tilde
A=\left(\begin{array}{@{}c@{}c@{}c@{\hspace{1mm}}cc@{\hspace{1.5mm}}c@{}} 
0 & 0 & 0 & 0 & 0 & 0\\  
0 & 0 & 0 & 0 & 0 & 0\\ 
u & 0 & 0 & 0 & 0 & 0\\ 
\tilde A_{41} & 0 & D_y U & 0 & 0 & -U\\ 
0 & 3V & 0 & 0 & 0 & 0\\
D_y U  & 0 & U & 0 & 0 & 0
\end{array}\right), \quad 
\tilde B=\left(\begin{array}{@{}c@{}c@{}cccc@{}}
0 & 0 & 0 & 0 & 0 & 0\\ 
U & 0 & 0 & 0 & 0 & 0\\ 
0 & 0 & 0 & 0 & 0 & 0\\ 
\tilde B_{41} & D_x U & 0 & 0 & -U & 0\\ 
D_x U & U & 0 & 0 & 0 & 0\\
0 & 0 & 3W & 0 & 0 & 0
\end{array}\right)
\]
are the 
directional derivatives of the matrices $A$ and $B$ (\ref{zcr}) along the
vector $(U,V,W)^T$, 
$\tilde A_{41}=-D_{y}^2(U)+3w U+ 3 u W$, 
$\tilde B_{41}=-D_{x}^2(U)+3v U+ 3u V$.
    

The next step is to find linear combinations $\tilde U,\tilde V,\tilde W$
of entries of
$S$ that solve (\ref{lin}).
A straightforward but tiresome computation shows that 
for $A$ and $B$ (\ref{zcr}) these are $\tilde
U=-S_{35}+S_{26}$,
$\tilde V=S_{54}+S_{12}$,
$\tilde W=-S_{13}-S_{64}$, i.e.
if $\bi{U}=(U,V,W)^{T}$
is a symmetry of (\ref{snv}), then
so is $\tilde\bi{U}=(\tilde U,\tilde V,\tilde W)^{T}$.
Hence, the linear operator $\mathfrak{R}_0$
mapping $\bi{U}$ to $\tilde\bi{U}$ 
is a recursion operator for (\ref{snv}).
\looseness=-1 

However, the application of 
$\mathfrak{R}_0$ to the simplest 
symmetries of (\ref{snv}), e.g. to the zero symmetry, yields nonlocal
symmetries of (\ref{snv}), so we should invert $\mathfrak{R}_0$ in order
to obtain a  
recursion operator
$\tilde\mathfrak{R}=\mathfrak{R}_0^{-1}$ generating hierarchies of local
symmetries for (\ref{snv}). 
\looseness=-1 

It turns out that 
our $\mathfrak{R}_0$
is invertible only for 
$\lambda\neq 0$.
Inverting $\mathfrak{R}_0$
involves solving a system
of algebraic and differential equations for the components of 
$\tilde\mathfrak{R}$, which is a fairly tiresome but 
algorithmic process.    
For the sake of simplicity we set all the integration constants 
to zero. Then
$\mathfrak{R}=\lambda\tilde\mathfrak{R}-\frac{1}{2}\lambda^2\id$, where
$\id$ is the identity operator, is independent of $\lambda$ and provides a 
conventional recursion operator for (\ref{snv}).
\looseness=-1 

The action of
$\mathfrak{R}$ on a symmetry $\bi{U}=(U,V,W)^T$ of (\ref{snv}) is given by
$\mathfrak{R}(\bi{U})=\mathfrak{L}(\bi{U})+\mathfrak{M}\vec Z$. Here   
$\vec Z=(Z_1,Z_2,Z_3,Z_4,Z_5)^{T}$ is a general solution of the system
\begin{equation}\label{cov}
\begin{array}{l}
\fl D_x Z_1=U, \ \ D_y Z_1=W, \ \
D_x Z_2 = V, \ \  D_y Z_2 =U,\\
\fl D_x Z_3 =D_y^2 U-3 (W u+ U w), \ \ 
D_y Z_3=D_x^2 U -3(V u+U v),\\
\fl D_x Z_4 =
-\frac{1}{3} v D_y^2 U
+ \frac{1}{3} u_x D_y U
+ (-\frac{1}{3} u_{xy} + u^2 + v w) U
+ (-\frac{1}{3} u_{yy} + \frac{1}{3} v_{xx}\\ + 2 u w - v^2) V
+ u v W
+ u u_x Z_1
+ (w u_x + u u_y - v v_x) Z_2,
\\
\fl D_y Z_4 =
-\frac{1}{3} v D_x^2 U 
+ \frac{1}{3} v_x D_x U 
+ (-\frac{1}{3} u_{yy} + 2 u w) U
+ u v V 
+ u u_y Z_1 
+ u^2 W\\ 
+ (-v u_x + w u_y + u w_y) Z_2,
\\
\fl D_x Z_5 =
-\frac{1}{3} w D_y^2 U 
+ \frac{1}{3} w_y D_y U 
+ (-\frac{1}{3} u_{xx} + 2 u v) U
+ u^2 V
+ u w W\\ 
+ (v u_x - w u_y + u v_x) Z_1 
+ u u_x Z_2,
\\
\fl D_y Z_5 =
-\frac{1}{3} w D_x^2 U 
+ \frac{1}{3} u_y D_x U 
+ (-\frac{1}{3} u_{xy} + u^2 + v w) U 
+ u w V\\
+ (-\frac{1}{3} u_{xx} + \frac{1}{3} w_{yy} + 2 u v - w^2) W 
+ (u u_x + v u_y - w w_y) Z_1 
+ u u_y Z_2.
\end{array}
\end{equation}
Note that this system is compatible if and only if $\bi{U}$ 
solves (\ref{lin}).

The operators $\mathfrak{L}$ and $\mathfrak{M}$ are of the form
\[
\mathfrak{L}=\left(\begin{array}{ccc}
\mathfrak{L}_{11} & \mathfrak{L}_{12} & \mathfrak{L}_{13}\\
\mathfrak{L}_{21} & \mathfrak{L}_{22} & \mathfrak{L}_{23}\\
\mathfrak{L}_{31} & \mathfrak{L}_{32} & \mathfrak{L}_{33}
\end{array}\right),\quad 
\mathfrak{M}=\left(\begin{array}{ccccc}
\mathfrak{M}_{11} & \mathfrak{M}_{12} 
& \mathfrak{M}_{13} & \mathfrak{M}_{14} 
& \mathfrak{M}_{15}\\
\mathfrak{M}_{21} & \mathfrak{M}_{22} 
& \mathfrak{M}_{23} & \mathfrak{M}_{24}
& \mathfrak{M}_{25}\\
\mathfrak{M}_{31} & \mathfrak{M}_{32} 
& \mathfrak{M}_{33} & \mathfrak{M}_{34} 
& \mathfrak{M}_{35}
\end{array}\right),  
\]
\[
\begin{array}{@{}l}
\fl \mathfrak{L}_{11}=D_{x}^6 - 6 v D_{x}^4 - \frac{25}{9} u
D_{x}^2D_{y}^2 - 15 v_x D_{x}^3 - \frac{2}{9} u_y D_{x}^2D_{y} -
\frac{29}{9} u_x D_{x}D_{y}^2 + (-\frac{5}{3} u_{yy} - 18 v_{xx}\\ +
\frac{40}{3} u w + 9 v^2) D_{x}^2 + 9 u^2 D_{x}D_{y} + (-\frac{5}{3}
u_{xx} + \frac{13}{3} u v) D_{y}^2 + (-3 u_{xyy} - 12 v_{xxx}\\ +
\frac{56}{3} w u_x + 26 u u_y + 27 v v_x) D_{x} + (26 u u_x + \frac{2}{3}
v u_y) D_{y} - 3 u_{xxyy} - 3 v_{xxxx}\\ + 14 w u_{xx} + 20 u u_{xy} +
5 v u_{yy} + 9 v v_{xx} + \frac{77}{3} u_x u_y + 9 v_x^2 - 4 u^3 - 28 u v
w,\\
\fl\mathfrak{L}_{12}=-\frac{28}{9} u D_{x}^4 - \frac{106}{9} u_x D_{x}^3
+ (-\frac{55}{3} u_{xx} + \frac{32}{3} u v) D_{x}^2 + (-\frac{44}{3}
u_{xxx} + \frac{74}{3} v u_x + 18 u v_x) D_{x}\\ 
- 6 u_{xxxx}
 + 19 v u_{xx} + 4 u u_{yy} + 10 u v_{xx} + \frac{79}{3} u_x v_x +
\frac{2}{3} u_y^2 - 12 u^2 w - 4 u v^2,\\ 
\fl\mathfrak{L}_{13}=-\frac{1}{9} u D_{y}^4 + \frac{2}{9} u_y D_{y}^3 +
(-\frac{1}{3} u_{yy} + \frac{5}{3} u w) D_{y}^2 + (\frac{1}{3} u_{xxx} -
v u_x - \frac{4}{3} w u_y - u v_x + u w_y) D_{y}\\ + 13 u u_{xx} + w
u_{yy} + u w_{yy} + \frac{29}{3} u_x^2 - \frac{2}{3} u_y w_y - 12 u^2 v -
4 u w^2,\\ 
\fl\mathfrak{L}_{21}=
\frac{28}{27} D_{x}^4 D_{y}^2 - \frac{28}{9} w D_{x}^4 - 6 u D_{x}^3D_{y}
- \frac{32}{9} v D_{x}^2 D_{y}^2 - \frac{86}{9} u_y D_{x}^3 -
\frac{134}{9} u_x D_{x}^2 D_{y} - \frac{58}{9} v_x D_{x}D_{y}^2\\ +
(-\frac{148}{9} u_{xy} + \frac{28}{3} u^2 + \frac{32}{3} v w) D_{x}^2 +
(-\frac{154}{9} u_{xx} + \frac{46}{3} u v) D_{x}D_{y} 
+ (-\frac{16}{9} u_{yy}\\ - \frac{28}{9} v_{xx} + \frac{16}{3} u w 
+ \frac{4}{3} v^2)
D_{y}^2 + (-14 u_{xxy} + \frac{89}{3} u u_x + \frac{50}{3} v u_y +
\frac{58}{3} w v_x) D_{x}\\ + (-\frac{86}{9} u_{xxx} 
+ \frac{58}{3} v u_x +
\frac{59}{3} u v_x) D_{y} - 6 u_{xxxy} + 20 u u_{xx} + \frac{40}{3} v
u_{xy} + \frac{16}{3} w u_{yy}\\ + \frac{28}{3} w v_{xx} + \frac{47}{3}
u_x^2 + 19 u_y v_x - 12 u^2 v - 16 u w^2 - 4 v^2 w,\\
\fl\mathfrak{L}_{22}=-\frac{1}{27} D_{x}^6 + \frac{2}{3} v D_{x}^4 + v_x
D_{x}^3 + (-\frac{28}{9} u_{yy} + \frac{11}{9} v_{xx} + \frac{28}{3} u w
- 3 v^2) D_{x}^2 + (-\frac{56}{9} u_{xyy}\\ + \frac{10}{9} v_{xxx} +
\frac{56}{3} w u_x + \frac{56}{3} u u_y - 7 v v_x) D_{x} - \frac{40}{9}
u_{xxyy} + \frac{5}{9} v_{xxxx} + \frac{40}{3} w u_{xx}\\ 
+\frac{52}{3} u u_{xy} + \frac{16}{3} v u_{yy} - \frac{16}{3} v v_{xx} +
\frac{70}{3} u_x u_y - 2 v_x^2 - 4 u^3 - 16 u v w + 4 v^3,\\
\fl\mathfrak{L}_{23}=-\frac{1}{9} u_x D_{y}^3 + (\frac{1}{3} u_{xy} -
\frac{1}{3} u^2) D_{y}^2 + (-\frac{2}{3} u_{xyy} + \frac{5}{3} w u_x + 2
u u_y) D_{y} - \frac{19}{9} u_{xxxx} + \frac{23}{3} v u_{xx} \\ 
- w u_{xy} + \frac{13}{3} u u_{yy} + \frac{19}{3} u v_{xx} + \frac{40}{3}
u_x v_x + \frac{4}{3} u_x w_y - 12 u^2 w - 4 u v^2,\\
\end{array}
\]
\[
\begin{array}{l}
\fl\mathfrak{L}_{31}=\frac{28}{27} D_{x}^5 D_{y} - 6 u D_{x}^4 -
\frac{56}{9} v D_{x}^3 D_{y} - \frac{4}{9} w D_{x}^2 D_{y}^2 -
\frac{142}{9} u_x D_{x}^3 + (-\frac{28}{3} v_x - \frac{2}{9} w_y)
D_{x}^2D_{y}\\ - \frac{22}{9} u_y D_{x}D_{y}^2 + (-\frac{184}{9} u_{xx} +
\frac{70}{3} u v + \frac{4}{3} w^2) D_{x}^2 + (-\frac{14}{9} u_{yy} -
\frac{28}{3} v_{xx} + \frac{20}{3} u w\\ + \frac{28}{3} v^2) D_{x} D_{y} +
(-\frac{8}{9} u_{xy} + \frac{2}{3} u^2 + \frac{4}{3} v w) D_{y}^2 +
(-\frac{142}{9} u_{xxx} + \frac{116}{3} v u_x + \frac{28}{3} w u_y\\ +
\frac{82}{3} u v_x + u w_y) D_{x} + (-\frac{14}{9} u_{xyy} - \frac{28}{9}
v_{xxx} + \frac{20}{3} w u_x + \frac{37}{3} u u_y + \frac{28}{3} v v_x\\ 
+ \frac{2}{3} v w_y) D_{y} - 6 u_{xxxx} + \frac{70}{3} v u_{xx} +
\frac{14}{3} w u_{xy} + \frac{8}{3} u u_{yy} + \frac{52}{3} u v_{xx} +
\frac{82}{3} u_x v_x\\ + \frac{1}{3} u_x w_y + 7 u_y^2 - 12 u^2 w - 16 u
v^2 - 4 v w^2,\\
\fl\mathfrak{L}_{32}=-\frac{29}{9} u_y D_{x}^3 + (-9 u_{xy} +
\frac{25}{3} u^2) D_{x}^2 + (-10 u_{xxy} + \frac{88}{3} u u_x + 11 v u_y)
D_{x} - \frac{47}{9} u_{xxxy}\\ + \frac{58}{3} u u_{xx} + \frac{44}{3}
v u_{xy} + \frac{4}{3} w u_{yy} + \frac{49}{3} u_x^2 + \frac{23}{3} u_y
v_x + \frac{2}{3} u_y w_y - 12 u^2 v - 4 u w^2,\\
\fl\mathfrak{L}_{33}=-\frac{1}{27} D_{y}^6 + \frac{2}{3} w D_{y}^4 + w_y
D_{y}^3 + (\frac{11}{9} w_{yy} - 3 w^2) D_{y}^2 + (\frac{10}{9} w_{yyy} -
7 w w_y) D_{y} - \frac{4}{3} u_{xxyy}\\ + \frac{5}{9} w_{yyyy} +
\frac{16}{3} w u_{xx} + \frac{26}{3} u u_{xy} + 4 v u_{yy} - \frac{16}{3}
w w_{yy} + \frac{40}{3} u_x u_y - 2 w_y^2\\ - 4 u^3 - 16 u v w + 4 w^3;\\
\fl\mathfrak{M}_{11}=-u_{xxxyy} + 2 w u_{xxx} + \frac{14}{3} u u_{xxy} +
3 v u_{xyy} + \frac{1}{3} u w_{yyy} + \frac{23}{3} u_y u_{xx} + 9 u_x
u_{xy} + 3 v_x u_{yy}\\ + w_y u_{yy} - \frac{2}{3} u_y w_{yy} - 4 u^2
u_x - 6 v w u_x - 8 u v u_y - 6 u w v_x - 4 u w w_y,\\
\fl\mathfrak{M}_{12}=-u_{xxxxx} + 5 v u_{xxx} + \frac{5}{3} u u_{xyy} +
\frac{10}{3} u v_{xxx} + 10 v_x u_{xx} + \frac{5}{3} u_x u_{yy} +
\frac{25}{3} u_x v_{xx} - 8 u w u_x \\ - 6 v^2 u_x - 4 u^2 u_y - 10 u v
v_x,\\ 
\fl\mathfrak{M}_{13}=-\frac{2}{3}u_{xxx}+2 v u_x+2 u v_x, \quad
\mathfrak{M}_{14}=-2 u_x,\quad \mathfrak{M}_{15}=-2 u_y,\\
\fl\mathfrak{M}_{21}=-u_{xxxxy} + 5 u u_{xxx} + 3 v u_{xxy} + 2 w u_{xyy}
+ \frac{32}{3} u_x u_{xx} + 6 v_x u_{xy} - w_y u_{xy} + u_y u_{yy}\\ + 3
u_y v_{xx} + \frac{1}{3} u_x w_{yy} - 9 u v u_x - 6 w^2 u_x - 9 u w u_y
- 5 u^2 v_x + u^2 w_y,\\
\fl\mathfrak{M}_{22}=-\frac{10}{9} u_{xxxyy} + \frac{1}{9} v_{xxxxx} +
\frac{10}{3} w u_{xxx} + 5 u u_{xxy} + \frac{10}{3} v u_{xyy} -
\frac{5}{3} v v_{xxx} + \frac{25}{3} u_y u_{xx} + 10 u_x u_{xy}\\ +
\frac{10}{3} v_x u_{yy} - \frac{5}{3} v_x v_{xx} - 10 v w u_x - 4 u^2 u_x
- 10 u v u_y + 4 v^2 v_x - 8 u w v_x,\\
\fl\mathfrak{M}_{23}=-\frac{2}{3}u_{xyy}+2 w u_x +2 u u_y, \quad
\mathfrak{M}_{24}= -2 v_x,\quad \mathfrak{M}_{25}=-2 u_x,\\ 
\fl\mathfrak{M}_{31}=-\frac{10}{9} u_{xxxxx} + \frac{1}{9} w_{yyyyy} +
\frac{20}{3} v u_{xxx} + \frac{5}{3} u u_{xyy} + \frac{10}{3} u v_{xxx} -
\frac{5}{3} w w_{yyy} + 10 v_x u_{xx} + \frac{5}{3} u_x u_{yy}\\ + 10 u_x
v_{xx} - \frac{5}{3} w_y w_{yy} - 10 v^2 u_x - 10 u w u_x - 4 u^2 u_y -
10 u v v_x + 4 w^2 w_y + 2 u v w_y,\\
\fl\mathfrak{M}_{32}=-u_{xxxxy} + 5 u u_{xxx} + 5 v u_{xxy} + 10 u_x
u_{xx} + 5 v_x u_{xy} + \frac{5}{3} u_y u_{yy} + \frac{10}{3} u_y v_{xx}\\
- 15 u v u_x - 6 v^2 u_y - 3 u w u_y - 5 u^2 v_x + u^2 w_y,\\ 
\fl\mathfrak{M}_{33}=-\frac{2}{3} u_{xxy} + 2 u u_x + 2 v u_y,\quad 
\mathfrak{M}_{34}=-2 u_y,\quad \mathfrak{M}_{35}=-2 w_y.
\end{array}
\]

Note that the above formula $\mathfrak{R}(\bi{U})=\mathfrak{L}(\bi{U})
+\mathfrak{M}\vec Z$
defines a recursion operator in the sense of Guthrie \cite{gut1},
and the system (\ref{cov}) defines a covering \cite{vinbook} over 
(\ref{lin}). 
Formally, we could express $Z_i$ from 
(\ref{cov}) as
$Z_1 = D_x^{-1} U, Z_2 = D_x^{-1} V$ etc., 
and thus write $\mathfrak{R}$ as an integro-differential operator,
as it became traditional in the literature, see, e.g.,
\cite{konbook, olv_eng2, bl}. However, if we drop the $y$-part of 
(\ref{cov}), we encounter certain difficulties in constructing new
symmetries, cf, e.g., \cite{gut1, sw1, serg}.
\looseness=-1

As integrating (\ref{cov}) involves arbitrary constants, we have
$\mathfrak{R}(0)=c_1 \bi{S}_1+c_2 \bi{S}_2+\cdots+c_5 \bi{S}_5$, 
where $c_i$ are 
constants, and $\bi{S}_{1},\dots, \bi{S}_{5}$ are symmetries of
(\ref{snv}) of the following form: 
\begin{eqnarray*}
\fl
\bi{S}_1=\!
\left(\begin{array}{@{}c@{}}
\mathfrak{M}_{11}+\frac12 u^2\mathfrak{M}_{14}
+(uv-\frac12 w^2)\mathfrak{M}_{15} \\
\mathfrak{M}_{21}+\frac12 u^2\mathfrak{M}_{24}
+(uv-\frac12 w^2)\mathfrak{M}_{25} \\
\mathfrak{M}_{31}+\frac12 u^2\mathfrak{M}_{34}
+(uv-\frac12 w^2)\mathfrak{M}_{35}
\end{array}\right)\!\!,
\ \
\bi{S}_2=\!
\left(\begin{array}{@{}c@{}}
\mathfrak{M}_{12}+(uw-\frac12 v^2)\mathfrak{M}_{14}
+\frac12 u^2\mathfrak{M}_{15} \\
\mathfrak{M}_{22}+(uw-\frac12 v^2)\mathfrak{M}_{24}
+\frac12 u^2\mathfrak{M}_{25} \\
\mathfrak{M}_{32}+(uw-\frac12 v^2)\mathfrak{M}_{34}
+\frac12 u^2\mathfrak{M}_{35}
\end{array}\right)\!\!,
\\
\fl
\bi{S}_3=
\left(\begin{array}{@{}c@{}}
-u_{xxx}+3(v u_x+u v_x) \\
-u_{xyy}+3(w u_x+u u_y) \\ 
-u_{xxy}+3(v u_y+ u u_x)
\end{array}\right)\!\!,
\quad
\bi{S}_4=\bi{u}_x\equiv \left(\begin{array}{@{}c@{}} u_x \\ v_x \\ w_x
\end{array}\right)\!\!,
\quad
\bi{S}_5=\bi{u}_y\equiv \left(\begin{array}{@{}c@{}} u_y \\ v_y \\ w_y
\end{array}\right).
\end{eqnarray*}

The repeated application of $\mathfrak{R}$ to $\bi{S}_{1},\dots,\bi{S}_5$
produces five hierarchies of symmetries of the stationary NVN equation 
(\ref{snv}), which can be visualized as follows
(numbers in the top line denote the orders of symmetries):
$$
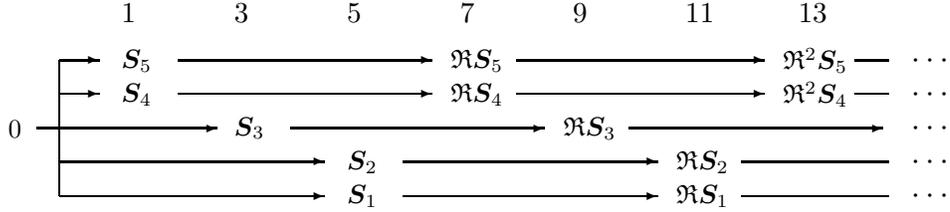
\unitlength=.75mm
\begin{picture}(150,35)
\put(10,32.5){\makebox(0,0)[l]{\small1}}
\put(30,32.5){\makebox(0,0)[l]{\small3}}
\put(50,32.5){\makebox(0,0)[l]{\small5}}
\put(70,32.5){\makebox(0,0)[l]{\small7}}
\put(90,32.5){\makebox(0,0)[l]{\small9}}
\put(110,32.5){\makebox(0,0)[l]{\small11}}
\put(130,32.5){\makebox(0,0)[l]{\small13}}
\put(-10,10.3){\footnotesize{$0$}}
\put(-1,24){\vector(1,0){7}}
\put(-1,18){\vector(1,0){7}}
\put(-5,12){\vector(1,0){32}}
\put(-1,6){\vector(1,0){47}}
\put(-1,0){\vector(1,0){47}}
\put(-1,0){\line(0,1){24}}
%
%
\put(10,24){\makebox(0,0)[l]{\footnotesize$\bi{S}_{5}$}}
\put(73,24){\makebox(0,0){\footnotesize$\mathfrak R \bi{S}_{5}$}}
\put(133,24){\makebox(0,0){\footnotesize$\mathfrak R^2 \bi{S}_{5}$}}
\put(20,24){\vector(1,0){45}}
\put(80,24){\vector(1,0){44}}
\put(140,24){\line(1,0){6}}
\put(150,24){\makebox(0,0)[l]{$\cdots$}}
\put(10,18){\makebox(0,0)[l]{\footnotesize$\bi{S}_{4}$}}
\put(73,18){\makebox(0,0){\footnotesize$\mathfrak R \bi{S}_{4}$}}
\put(133,18){\makebox(0,0){\footnotesize$\mathfrak R^2 \bi{S}_{4}$}}
\put(20,18){\vector(1,0){45}}
\put(80,18){\vector(1,0){44}}
\put(140,18){\line(1,0){6}}
\put(150,18){\makebox(0,0)[l]{$\cdots$}}
\put(30,12){\makebox(0,0)[l]{\footnotesize$\bi{S}_{3}$}}
\put(93,12){\makebox(0,0){\footnotesize$\mathfrak R \bi{S}_{3}$}}
\put(40,12){\vector(1,0){45}}
\put(100,12){\vector(1,0){45}}
\put(150,12){\makebox(0,0)[l]{$\cdots$}}
\put(50,6){\makebox(0,0)[l]{\footnotesize$\bi{S}_{2}$}}
\put(113,6){\makebox(0,0){\footnotesize$\mathfrak R \bi{S}_{2}$}}
\put(60,6){\vector(1,0){45}}
\put(120,6){\line(1,0){26}}
\put(150,6){\makebox(0,0)[l]{$\cdots$}}
\put(50,0){\makebox(0,0)[l]{\footnotesize$\bi{S}_{1}$}}
\put(113,0){\makebox(0,0){\footnotesize$\mathfrak R \bi{S}_{1}$}}
\put(60,0){\vector(1,0){45}}
\put(120,0){\line(1,0){26}}
\put(150,0){\makebox(0,0)[l]{$\cdots$}}
\end{picture}
$$
We conjecture that all these symmetries are local and commute, 
as it is the case for
the symmetries of orders $1,3,5,\dots,11$.
\looseness=-1

Note that (\ref{snv}) has a scaling symmetry 
$\bi{S}=x\bi{u}_x+y\bi{u}_y+2 \bi{u}$. 
The application of $\mathfrak{R}$ to $\bi{S}$ yields a 
nonlocal  symmetry of seventh order, which we conjecture to be a master
symmetry for (\ref{snv}),  meaning that commuting $\mathfrak{R}(\bi{S})$
with any symmetry belonging to  one of the five hierarchies, described 
above, yields (up to a constant  multiplier) the next member of the same
hierarchy.  The repeated application of $\mathfrak{R}$ to $\bi{S}$ yields
an infinite  hierarchy of nonlocal symmetries for (\ref{snv}).

We believe that $\mathfrak{R}$ is hereditary
in the sense of \cite{ff}, but we have not yet checked 
this because of the huge amount of
computations involved.  \looseness=-1

As a final remark, let us mention the nonstandard structure 
of nonlocal terms of 
$\mathfrak{R}$ in (\ref{cov}): they involve the derivatives 
of components of the symmetry,
what is quite unusual, cf, e.g., \cite{sak} for another example of this kind
and \cite{wang} for a comprehensive list of known today integrable systems
in (1+1) dimensions and their 
recursion operators. 


\subsection*{Acknowledgements}
We are sincerely grateful to Dr E V Ferapontov and Dr M V Pavlov
for stimulating discussions.
This research was supported in part by the
Ministry of Education, Youth and Sports of Czech Republic under grant
MSM:J10/98:192400002 (MM \& AS),  and by the Czech Grant Agency under
grant no 201/00/0724 (AS).

\let\section\subsection

\end{document}